\documentclass[aps,prl,twocolumn,showpacs,amsmath,amssymb]{revtex4}
\makeatletter
\def\@dotsep{4.5}
\makeatother

\usepackage{graphicx}

\newlength{\onefig}
\newlength{\twofig}
\newcommand{\bq}{\begin{eqnarray}}
\newcommand{\eq}{\end{eqnarray}}
\newcommand{\bqn}{\begin{eqnarray*}}
\newcommand{\eqn}{\end{eqnarray*}}
\newcommand{\beq}{\begin{equation}}
\newcommand{\eeq}{\end{equation}}

\newcommand{\ve}[1]{\mathbf{#1}}

\newcommand{\etal}{\textit{et~al.}}

\begin{document}

\title{Single-particle and collective slow dynamics of colloids in porous confinement}

\author{Jan Kurzidim}
\author{Daniele Coslovich}
\author{Gerhard Kahl}
\affiliation{Insitut f\"ur Theoretische Physik and CMS, Technische Universit\"at
  Wien -- Wiedner Hauptstra{\ss}e 8-10, A-1040 Wien, Austria} 

\date{\today}

\begin{abstract}
Using molecular dynamics simulations we study the slow dynamics of a hard sphere fluid confined in a disordered porous matrix. The presence of both discontinuous and continuous glass transitions as well as the complex interplay between single-particle and collective dynamics are well captured by a recent extension of mode-coupling theory for fluids in porous media. The degree of universality of the mode-coupling theory predictions for related models of colloids is studied by introducing size-disparity between fluid and matrix particles, as well as softness in the interactions.
\end{abstract}


\maketitle

Investigations of the dynamic properties of fluids confined in a
disordered porous matrix have recently attracted considerable
attention in the scientific community both from a fundamental and from
a technological point of view. Of particular interest is the glass
formation process where the single-particle and the collective
dynamics of the fluid are exposed to the complex interplay between
confinement and connectivity of the pores. While a considerable amount
of experimental results has been compiled in this
field~\cite{alcoutlabi_effects_2005,mckenna_confit_2007}, theoretical
investigations are still rare: remarkable computer simulations
on a few selected state points and systems have been
performed~\cite{kim_effects_2003,gallo_slow_2003}, but systematic
studies of this phenomenon are difficult due to
the large computational cost caused by the requirement of long
simulations and by the thermodynamic averaging prescriptions
for these systems.

A breakthrough in the theoretical investigations was achieved with the
formalism put forward by
Krakoviack~\cite{krakoviack_liquid-glass_2005,%
  krakoviack_mode-coupling_2007,krakoviack_tagged-particle_2009} who
successfully combined two concepts to study the 
dynamics of
fluids confined in disordered porous matrices. On one hand, his
framework is based on the replica Ornstein-Zernike (ROZ)
formalism~\cite{given_comment:_1992}, where the system at hand is
viewed as a quenched-annealed (QA) mixture of mobile fluid particles
and immobile matrix particles. The other ingredient is mode-coupling
theory (MCT)~\cite{gtze_relaxation_1992}: based on the {\it static}
correlations of the system,
MCT predicts the time dependence of the single-particle
(``self'') and collective density correlators.
MCT has been able to identify new and unexpected features in the
dynamics of colloidal
systems~\cite{sciortino_glassy_2005,zaccarelli_colloidal_2007,mayer_asymmetric_2008,sciortino_evidence_2003}
and provides a convincing account of the early stages of the
structural arrest in molecular glass-forming
liquids~\cite{gtze_recent_1999}.

The results for the kinetic diagram of a simple hard sphere (HS) fluid
in a HS matrix, evaluated using the MCT framework with ROZ structure
factors as an input, contains indeed a wealth of intriguing
features~\cite{krakoviack_mode-coupling_2007,krakoviack_tagged-particle_2009}:
(i) two types of glass transitions are encountered, a discontinuous
type B transition at low matrix packing fractions ($\phi_m$), and a
continuous type A transition at large $\phi_m$~\footnote{%
The infinite time limit of density correlators changes discontinuously (continuously) from zero to finite values at a type B (type A) transition.}; 
the two transition
lines meet at a degenerate high order singularity point
(A$_3$). Further, the theory predicts (ii) a re-entrant glass
transition for large $\phi_m$ and (iii) a continuous
diffusion-localization transition that occurs only in the self
dynamics; the latter transition is driven by the localization of
particles in disconnected void domains formed by the
matrix~\cite{krakoviack_tagged-particle_2009}, as in the Lorentz
model~\cite{hfling_localization_2006}.

This Letter aims in two directions: on one side we provide for the
first time a parallel study to Krakoviack's main theoretical
predictions by means of computer simulations for the same HS
system. The simulations provide, in addition, a ``realistic'' picture
of the above dynamic effects as opposed to the ``idealized'' view that
emerges from MCT. On the other hand, with the help of the ROZ+MCT
framework we explore the parameter space of two further types of
potentials to find out to which degree the predicted dynamic features
are generic.  To this end we mimic typical interactions encountered in
colloidal systems by introducing a size disparity between matrix and
fluid particles (establishing thereby a link to the size-asymmetric
mixtures investigated
in~\cite{thakur_glass_1991,moreno_anomalous_2006,voigtmann_double_transition_2008})
and softness in the interparticle interaction.

By means of event-driven molecular dynamics (MD) we study a QA system of
hard spheres in three dimensions under periodic boundary
conditions. As in the monodisperse model studied
in~\cite{krakoviack_mode-coupling_2007}, both matrix and fluid
particles have the same diameter
$\sigma_{m}=\sigma_{f}=\sigma$ 
and mass $m$. The
total number of particles was $N=N_f+N_m=1000$. In the following, we
use $\sigma$ and $\sqrt{k_B T/m\sigma^2}$ as units of length and time,
respectively~\cite{allen_computer_1987}. The porous matrix
was generated by quenching an equilibrium fluid configuration
at a packing fraction
$\phi_m=\pi/6\rho_m\sigma_{m}^3$. Subsequently, the fluid component
with packing fraction $\phi_f=\pi/6\rho_f\sigma_{f}^3$ was
inserted into the remaining volume using an
optimized inflation algorithm.
For each state point, dynamic properties were averaged over
ten independent realizations of the system.
The fluid component was considered equilibrated
if the total mean squared displacement (MSD) exceeded $10^2$ within
a simulation time of $3\times 10^4$. State points for
which at least half of the system realizations did not meet this criterion
were considered glassy. 
No signs of crystallization were observed.
The MCT equations for QA systems were
solved using the numerical procedure outlined
in~\cite{krakoviack_mode-coupling_2007,krakoviack_tagged-particle_2009}. For the HS system, the
connected and blocked structure factors, used as an input for the MCT equations, were determined by solving the ROZ
equations~\cite{given_comment:_1992} with the Percus-Yevick (PY)
closure relation. The so-obtained critical packing fractions and
exponents typically agree within few percent with those reported
in~\cite{krakoviack_mode-coupling_2007,krakoviack_tagged-particle_2009}.

\begin{figure}[tb]
\includegraphics*[width=\onefig]{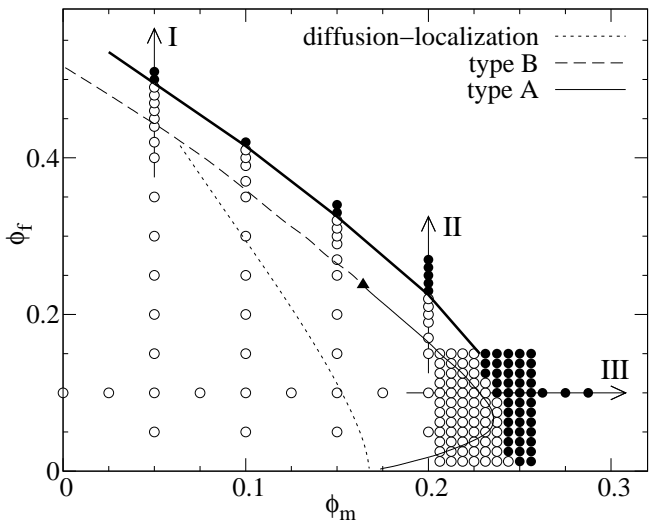} 
\caption{\label{fig:kinetic_diagram} Kinetic diagram of the QA
  HS system: MCT transition lines (thin lines; from
  Ref.~\cite{krakoviack_tagged-particle_2009}) and dynamic arrest line
  from simulations (thick line; see text for definition). The A$_3$
  singularity predicted by MCT is indicated by a filled triangle. Empty
  and filled circles indicate equilibrated and glassy simulated state
  points, respectively.}
\end{figure}
 
In Fig.~\ref{fig:kinetic_diagram} we show in the $(\phi_m,\phi_f)$-plane
the glass and the diffusion-localization transition lines obtained
from
MCT~\cite{krakoviack_mode-coupling_2007,krakoviack_tagged-particle_2009}
together with the dynamic arrest line determined from simulations. The
arrest line interpolates through points
with adjacent equilibrated and glassy state points. 
The shape of the arrest line resembles qualitatively the ideal MCT glass
line, but it starts bending downwards
more rapidly for $\phi_m\agt 0.2$.
Performing simulations on a fine mesh in
$\phi_f$ and $\phi_m$ we found that the arrest line is not re-entrant
at small $\phi_f$. We will show below that this apparent
discrepancy can be rationalized in terms of a crossover from glass
to diffusion-localization transition in the self dynamics.

\begin{figure}[tb]
\includegraphics*[width=\onefig]{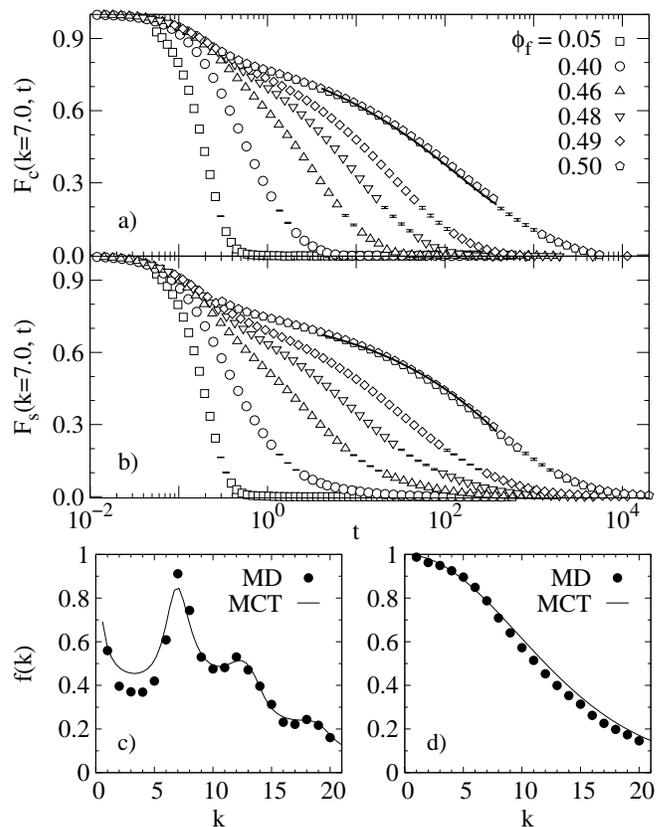} 
\caption{\label{fig:etam005} Upper panels: intermediate scattering
  functions at $\phi_m=0.05$ for different values of $\phi_f$: (a)
  connected part $F_c(k=7.0,t)$ and (b) self part
  $F_s(k=7.0,t)$. Solid lines are fits to the second-order expansion
  of the $\beta$-correlator (see text). 
  Error bars for selected times represent one standard deviation on the average over matrix realizations.
  Lower panels: non-ergodicity
  parameter $f(k)$ (symbols) for (c) connected and (d) self
  correlators at $\phi_m=0.05$, $\phi_f=0.50$. The corresponding
  non-ergodicity parameters predicted by MCT at $\phi_f=0.446$ are
  included as solid lines.}
\end{figure}

To investigate in more detail the different transition scenarios
predicted by MCT, we focus on selected paths (indicated by arrows and
labelled in Fig.~\ref{fig:kinetic_diagram}) across the
$(\phi_m,\phi_f)$-plane. We study both the self part
$F_s(k,t)=1/N\langle \sum_j \exp\{i \ve{k} \cdot
[\ve{r}_j(t)-\ve{r}_j(0)]\}\rangle$ and the connected part
$F_c(k,t)=\langle
\delta\rho_{\ve{k}}(t)\delta\rho_{\ve{k}}^*(0)\rangle/\langle\rho_{\ve{k}}(0)\rho_{\ve{k}}^*(0)\rangle$
of the intermediate scattering
function~\cite{krakoviack_mode-coupling_2007}, where
$\delta\rho_{\ve{k}}(t)=\rho_{\ve{k}}(t)-\langle\rho_{\ve{k}}\rangle$,
$\rho_{\ve{k}}(t)=\sum_j \exp[i\ve{k}\cdot \ve{r}_j(t)]$, and
$\langle\dots\rangle$ indicates a thermal average. $F_c(k,t)$ is the
appropriate correlator to characterize the slow collective dynamics of
fluids in porous media, since the conventional coherent intermediate
scattering function, $F(k,t)$, fails to decay to zero at long times
due to matrix-induced average density
fluctuations~\cite{krakoviack_mode-coupling_2007}.

Figure~\ref{fig:etam005} shows the evolution of the intermediate
scattering functions upon increasing $\phi_f$ at fixed $\phi_m=0.05$
(path I). The wave-vector considered is $k=7.0$, close to the main
peak of the fluid-fluid static structure factor. Both the self and the
collective dynamics slow down significantly as $\phi_f$ increases
towards 0.51, while a plateau at intermediate $t$ develops both in
$F_s(k,t)$ and in $F_c(k,t)$. These features are similar to those
observed in bulk glass-forming liquids and are compatible with a
``smeared'' type B transition. The difference between the MCT
critical packing fraction, $\phi_f^c=0.439645$, and the glass transition
observed in our simulations amounts to $\approx$ 15\%. Similar
discrepancies have been reported in investigations on bulk hard
spheres and suspensions of hard colloids~\cite{gtze_recent_1999}.

To test the predictions of MCT at a more quantitative level, we
describe the dynamics within the $\beta$-relaxation regime (close
to the plateau) using the second-order asymptotic expansion of the MCT
$\beta$-correlator,
$f(k)+H_1(k)t^b+H_2(k)t^{2b}$~\cite{gtze_relaxation_1992}. We perform
a simultaneous fit to all integer wave-vectors $1\leq k\leq 20$
at the largest equilibrated packing fraction, $\phi_f=0.50$. The
result for $k=7.0$ is included in Fig.~\ref{fig:etam005}; the exponent
$b$ is $0.33\pm 0.05$ for $F_s(k,t)$ and $0.36\pm 0.05$ for
$F_c(k,t)$. The range over which the fit can be considered reliable
covers $\approx 1.5$ decades in time. The estimated uncertainty on
$b$ accounts for the sensitivity of the fit to the choice of
the time range. The fitted values are somewhat smaller than $b=0.51$
(independent of correlators) obtained from numerical solution of
the MCT equations, which may be due to the limited time window
available for fitting the simulation data. Further, in the
bottom panels of Fig.~\ref{fig:etam005} we show the non-ergodicity parameter
$f(k)$ for both self and connected correlators. In both cases, $f(k)$
compares well with the MCT predictions at $\phi_f=0.446$, i.e.,
only slightly above $\phi_f^c$. Comparisons of similar quality have been
reported for bulk
glass-formers~\cite{nauroth_quantitative_1997,sciortino_debye-waller_2001}.
We conclude that MCT is able to describe at a semi-quantitative level
the slow dynamics of the QA system in this portion of the kinetic
diagram.

\begin{figure}[tb]
\includegraphics*[width=\onefig]{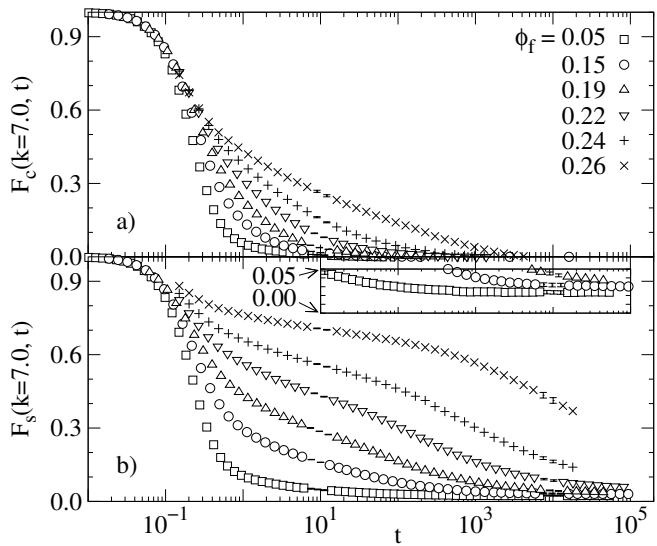} 
\caption{\label{fig:etam020} Intermediate scattering functions at
  $\phi_m=0.20$ for different values of $\phi_f$: (a) connected part
  $F_c(k=7.0,t)$ and (b) self part $F_s(k=7.0,t)$. 
  Error bars as in Fig.~\ref{fig:etam005}.  
  Inset of (b): enlarged view of the plateau of $F_s(k=7.0,t)$ at long times. }
\end{figure}

A more complex scenario appears at large $\phi_m$, where the stronger
influence of the matrix structure changes the nature of the transition. In
Fig.~\ref{fig:etam020} we show the intermediate scattering functions
for $k=7.0$ at constant $\phi_m=0.2$ (path II in
Fig.~\ref{fig:kinetic_diagram}). Along this path, the relaxation
patterns for self and collective correlators differ markedly as $\phi_f$ is varied. The
relaxation of $F_c(k,t)$ becomes slower as $\phi_f$ 
increases. Contrary to the case of dilute matrices, however, there is
no finite-height plateau at intermediate times. This indicates the
approach to a continuous type A glass transition, consistent with MCT
predictions along path II. On the other hand, the decay of $F_s(k,t)$
occurs in two steps, with a first inflection ($1<t<10$) and a
subsequent stretched decay at long times to a \textit{finite} plateau.
Upon increasing $\phi_f$ the plateau rises continuously starting from values
close to 0, as one expects in a type A transition. 
These features of $F_s(k,t)$ are consistent with those predicted by MCT across the continuous
diffusion-localization transition (dotted line in
Fig.~\ref{fig:kinetic_diagram}).

The cascade of decays in $F_s(k,t)$ hence arises from
the interplay between two dynamic effects: a weak, collective mechanism
of caging by fluid particles and an effective trapping in the voids
left by the matrix particles.
The superposition of glass and diffusion-localization transitions
leads to an effective ``decoupling'' between self and collective
dynamics. In fact, at the largest equilibrated $\phi_f$ 
along path II ($\phi_f=0.22$) the ratio $\tau_s(k)/\tau_c(k)$ of the relaxation times, defined via
the decay to $0.1$ of the corresponding correlators, is already larger
than $10^2$ for $k=7.0$. As a result, the dynamic arrest line (see
Fig.~\ref{fig:kinetic_diagram}) bends downwards rapidly in the
large-$\phi_m$, small-$\phi_f$ part of the diagram, following the
trend of the diffusion-localization line. 
Further work is required to assess
whether the predicted re-entrance of
the glass line exists or if it is hindered by the
progressive arrest of the self dynamics. 

\begin{figure}[tb]
\includegraphics*[width=\onefig]{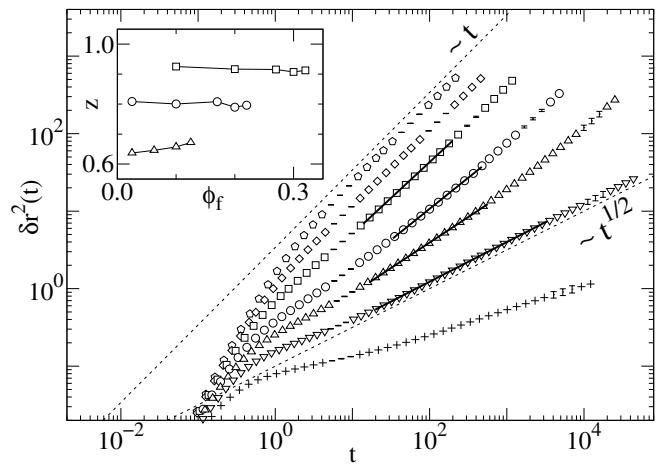} 
\caption{\label{fig:msd} MSD at constant
  $\phi_f=0.1$ for $\phi_m=0.05$, 0.10, 0.15, 0.20, 0.225, 0.25, and 0.275
  (from left to right). Solid lines are fits to $\delta r^2(t)
  \sim t^z$ in the sub-diffusive regime. 
  Error bars as in Fig.~\ref{fig:etam005}. 
  Inset: $z$ as a function of
  $\phi_f$ for selected $\phi_m$ (symbols as in the main plot).}
\end{figure}

According to MCT, the stretched decay of self correlators for large
$t$ close to the diffusion-localization line is associated with a
sub-diffusive behavior in the MSD $\delta
r^2(t)$~\cite{krakoviack_tagged-particle_2009}. In Fig.~\ref{fig:msd}
we show the evolution of the MSD at constant $\phi_f=0.10$ (path III
in Fig.~\ref{fig:kinetic_diagram}). Note that upon increasing $\phi_m$
along this path, the slowing down of the self dynamics is 
dictated by the diffusion-localization transition. For $\phi_m\alt 0.15$,
the ballistic regime $\delta r^2(t)\sim t^2$ is followed by
normal diffusion $\delta r^2(t)\sim t$. However, for $\phi_m \agt 0.15$,
the system becomes sub-diffusive, i.e., $\delta r^2(t)\sim t^z$ with
$z<1$, over an intermediate time window that becomes broader as
$\phi_m$ increases. As in the Lorentz
model~\cite{hfling_localization_2006}, $z$ decreases as $\phi_m$
increases. Up to a $\phi_m^*\approx 0.25$, normal diffusion is eventually
recovered at longer times, while for larger $\phi_m$ the MSD tends to
saturate. Close to $\phi_m^*$, the sub-diffusion
exponent is $z\approx 0.51$, in striking agreement with the MCT
prediction along the diffusion-localization line
($z=0.5$~\cite{krakoviack_tagged-particle_2009}). Remarkably, we also find that $z$ is nearly
independent of $\phi_f$ at fixed $\phi_m$ (see inset of
Fig.~\ref{fig:msd}). Hence, even at finite $\phi_f$, the
diffusion-localization transition is intimately related to the
geometrical properties of the matrix.

\begin{figure}[tb]
\includegraphics*[width=\onefig]{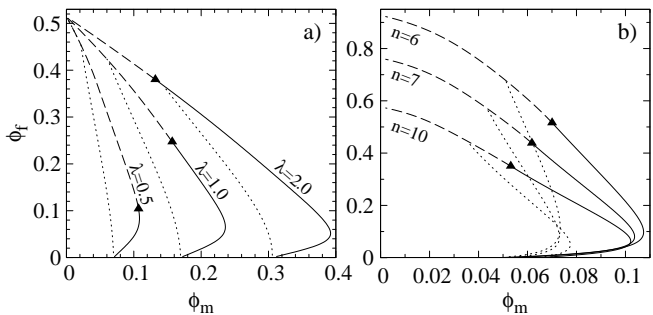}
\caption{\label{fig:others} MCT kinetic diagram for (a) QA hard
  spheres with size ratio $\lambda=0.5,1.0,2.0$ (from left to right)
  and (b) QA soft spheres with $\lambda=1.0$ and repulsive exponent
  $n=6,7,10$ (from top to bottom). Lines as in
  Fig.\ref{fig:kinetic_diagram}. Filled triangles indicate points where type
  A and B transitions meet.}
\end{figure}

To assess the degree of universality of the scenario
predicted by the ROZ+MCT formalism, we solved the MCT equations for
related QA systems based on purely repulsive interactions.
Specifically, we considered additive hard spheres with size ratio
$\lambda=\sigma_{m}/\sigma_{f}\neq 1$ and soft spheres interacting
with inverse power potentials $u(r)=\epsilon (\sigma/r)^n$ where $6\le
n\le 12$. In the latter case, we used $\lambda=1$ and the structure
factors were calculated at a temperature $T/\epsilon=0.2$ by solving
the ROZ equations in combination with the hypernetted-chain
approximation~\footnote{In the same range of $n$, the PY closure did not to lead to the correct variation 
of $\phi_f^c$ as a function of $n$ for $\phi_m\rightarrow 0$.}.
In Fig.~\ref{fig:others} we show results for selected
values of $\lambda$ and $n$. The topology of the MCT kinetic diagrams
of the studied systems is very similar to the one shown in
Fig.~\ref{fig:kinetic_diagram}, but subtle differences are observed as
system parameters vary. As $\lambda$ decreases in the HS systems, the
glass line shifts to smaller $\phi_m$ and the crossing
point between type B and A transitions moves towards the turning point
of the glass line. On the other hand, in the soft sphere systems the
glass line shifts to larger $\phi_f$ as $n$ decreases (as expected),
and the diffusion-localization line becomes clearly re-entrant. We plan
to investigate these features in more detail in future work.

In conclusion, we revealed the existence of a subtle interplay
between self and collective dynamics in a QA model of fluid in porous
confinement. The MCT framework for QA
systems~\cite{krakoviack_liquid-glass_2005,krakoviack_mode-coupling_2007,krakoviack_tagged-particle_2009}
provides idealized but consistent guidelines to explain the dynamic
features apparent for both dilute and dense matrices.
This
includes the superposition of glass transitions, driven by a
collective caging mechanism, and diffusion-localization transitions,
associated with the ceasing percolation of voids in the matrix
structure. The predicted features should be generic for a broad class
of model colloidal fluids adsorbed in porous media, and should be
observed experimentally in colloidal suspensions confined in porous
matrix configurations, quenched for instance by optical tweezers.

\begin{acknowledgements} 
The authors gratefully acknowledge computational aid by D.~Schwanzer
and helpful discussions with V.~Krakoviack. This work was funded by
the Austrian Science Foundation (FWF) under Proj. No. P19890-N16.
\end{acknowledgements}

\textit{Note}---While finalizing this paper, we became aware of a recent
paper by Kim~\etal~\cite{kim_slow_2009}, which focuses on a
related model. Their results for the HS QA
system are consistent with the analysis presented here.


\end{document}